# BeeTS: Smart Distributed Sensor Tuple Spaces combined with Agents using Bluetooth and IP Broadcasting

Stefan Bosse

University of Bremen, Dept. Mathematics & Computer Science, Bremen & Institute for Digitization, Bremen, Germany

**Abstract**. Most Internet-of-Things (IoT) devices and smart sensors are connected via the Internet using IP communication directly accessed by a server that collect sensor information periodically or event-based. Although, Internet access is widely available, there are places that are not covered and WLAN and mobile cell communication requires a descent amount of power not always available. Finally, the spatial context (the environment in which the sensor or devices is situated) is not considered (or lost) by Internet connectivity. In this work, smart devices communicate connectionless and ad-hoc by using low-energy Bluetooth broadcasting available in any smartphone and in most embedded computers, e.g., the Raspberry PI devices. Bi-directional connectionless communication is established via the advertisements and scanning modes. The communication nodes can exchange data via functional tuples using a tuple space service on each node. Tuple space access is performed by simple evenat-based agents. Mobile devices act as tuple carriers that can carry tuples between different locations. Additionally, UDP-based Intranet communication can be used to access tuple spaces on a wider spatial range. The Bluetooth Low Energy Tuple Space (BeeTS) service enables opportunistic, ad-hoc and loosely coupled device communication with a spatial context.

## Contents

Stefan Bosse    Smart Distributed Sensor Tuple Spaces combined with Agents

## 1.  Introduction

The number of deployed embedded systems increases exponentially. Ubiquitous and pervasive computing introduced a lot of visible and non-visible low-resource and mobile devices. Most Internet-of-Things (IoT) devices and smart sensors are still connected via the Internet using IP communication and that are accessed by a server that collect sensor information periodically or event-based. Although, Internet access is widely available, there are places that are not covered and WLAN and mobile cell communication requires a descent amount of power not always available. Additionally, the residential time of mobile devices can be below one Minute, not suitable for ad-hoc connection-based and negotiated communication. Finally, the spatial context (the environment in which the sensor or devices are situated) is not considered (or lost) by Internet connectivity. In this work, smart devices communicate connectionless and ad-hoc by using low-energy Bluetooth available in any Smartphone and in most embedded computers, e.g., the Raspberry PI devices. Bi-directional connectionless communication is established via the advertisement and scanning modes. The communication nodes can exchanged small data or functional tuples using a tuple space service. Mobile devices act as tuple carriers that can carry tuples between different locations. Additionally, UDP-based Intranet communication can be used to connect tuple spaces.

Tuple spaces are widely used for data storage for multi-processing in distributed and parallel systems. The Bluetooth Low Energy Tuple Space (BeeTS) service is a lazy distributed tuple space server and client. BeeTS uses Bluetooth and UDP broadcasting for tuple space interaction and tuple exchange. BeeTS supports tuples with an arity up to 4.

A tuple space provides a spatial context, i.e., tuple space access (by mobile devices) is limited to nearby devices. Distributed tuple spaces can be connected by node routers using IP communication if available. The router composes global space sets by tuple exchange and replication. The router is customised by function code rules. These rules can be changed at run-time and the code can use Machine Learning algorithms to optimally distribute tuples.

The novelty of this work is two-folded. Firstly, an ubiquitous radio broadcast medium is used for low-distance communication in ad-hoc mobile networks combined with a unified tuple space paradigm. Secondly, the tuple space communication is performed by simple reactive event-based agents programmed in JavaScript that can be sent to a node via the tuple space operations, too.

## 2.  Related Work

Originally, Bluetooth was introduced as a short-range wireless communication tech-nology used for linking peripherals (like ear phones) to smartphones. Mid-range and connectionless Bluetooth communication is used in several crowd sensing and crowd interaction use-cases, e.g., used for attendance tracking [1] and most prominently for pandemic contact tracing [2,3],[4]. These use-cases basically perform unidirectional producer-consumer communication using advertisement packet broadcasts. In this work, asynchronous bi-directional communication between nodes of a set that are within the receiving range is addressed.

In [5], multi-hop networks are established with BLE using a new multi-hop GATT layer service and connection-based node group connections (a group is a set of nodes within a specific spatial diameter). A connection protocol starting with advertisement, negotiation (requiring authorisation credentials), and final set-up of a bi-directional communication channel is time consuming and can require up to one second.

Instead using connection-based channel communication that is mostly not suitable to transfer small amounts of data, especially, from mobile devices with high spatial fluctuation, abusing the advertisement mode to transfer small amount of data with broadcasting is attractive. Small sensor nodes and economy advertisement tracking uses this kind of communication. Current deployment of smartphones for social contact tracing relies mostly on this method, too. But transferring data payload via advertisement packets has some drawback, mostly the issue of a low reception probability, discussed [6]. If the number of devices sending advertisement packets in the receiving range of a device increases, the probability $p_1$ to receive at least one advertisement packet from a sequence of identical packets decreases significantly. But nonetheless, [6] concluded that BLE broadcasting is still suitable for most situations.

The generative tuple space paradigm is well suited for ad-hoc mobile networks [7], especially if this paradigm is coupled with the agent paradigm [8]. The concept of fusion of mobile devices with a set of heterogenous sensors providing a sensor service that can be accessed wia the Internet was introduced by [9].

## 3. Formalisation of Distributed Tuple Spaces

A tuple space is basically a data base containing tuples. A tuple is a poly-typed ordered set of data values. The number of data values specifies the arity of the tuple, i.e., $tu=\langle v_1, v_2,..., v_k\rangle$, $|tu|=k$. Each tuple space *TS* can be divided into independent sub-spaces $TS= ts_1 \cup ts_2 \cup .. \cup ts_n$, with $ts_i$ holding only tuples of arity *i*. The data type of each element of a tuple can be arbitrary, i.e., any scalar value (float, integer, Boolean), or composed values, i.e., arrays or structures. In this work, the data elements are limited to scalar values, more specifically, float32, int16, and short strings (or data bytes).

Tuple space communication is generative, i.e., the lifetime of a tuple can be longer than the lifetime (or presence) of the producer process. There are producer and consumer processes. A producer process uses the *out(⟨tuple⟩)* operation to store tuple in the space. The consumer process uses the input operation *inp*(⟨*pattern*⟩) to remove a tuple. A tuple is found by pattern matching. A pattern is a tuple with actual and formal parameters (wild cards). Any matching tuple is returned. The input operation destroys the tuple atomically, i.e., one input request consumes at most one tuple. In distributed asynchronous systems this is difficult and expensive to be achieved. There is a read operation *rd*(⟨*pattern*⟩) that returns only a copy of the matching tuple. Tuples can be deleted by using the remove operation *rm*(⟨*tuple*|*pattern*⟩). Tuples can have an unlimited lifetime, but practically the lifetime is limited either by the tuple space services (removing old tuples) or by the *out* operation providing a lifetime $t$, i.e., $tu(t) = \langle v_1, v_2, .., v_k, t \rangle$, $|tu|=k$. The determination of the tuple lifetime is difficult in advance and depend on the application and the producer-consumer interaction time scale, but upper boundaries can be defined.

The read and input operations are typically synchronous, i.e., as long as there is no matching tuple the requesting process is blocked. This operational semantic requires in distributed systems reliable and synchronous bi-directional communication that is not available in this work. For this reason, the read operation is just a request that can be fulfilled within a time interval $[t_0, t_1]$, or not (time-out).

## 4. Unreliable Broadcast Communication

It is assumed that there is a broadcast medium *B*, e.g., using radio waves, which can reach a number of nodes $N_B = \{n_i\}_{i=1}^k$ defining a receive area/range coverage $C(B,N)(t)$ that changes with time $t$. *B* can send broadcast messages *m* to all listening nodes reachable by *B*. The set of nodes within *B* can vary on time and spatial scale. Furthermore, it is assumed that there is a probability $p_i(m, r_{i,j}, [t_0, t_1]) \in [0,1]$ that a message *m* is received by a node *i* sent by node *j* in distance *r* within a time interval $[t_0, t_1]$. These two assumptions are fundamental for the proposed distributed tuple spaces.

It is assumed that single packets that can be send over *B* are strictly limited by a small number of bits, e.g., 200-300 Bits. This requires a compact and optimized message format, discussed in the next sub-section.

### 4.1 Messages

There are seven different message types:

- *OUT* stores a tuple in all tuple spaces receiving this message;
- *RD* and *INP* requesting tuples from all receiving tuple spaces;
- *TEST* checks for the existence of a tuple or set of tuples;
- *TUPLE* is either an initial message sending this tuple to all receiving nodes without; storing the tuple in the respective tuple space, or a reply to a tuple request;

- *IAMHERE* and *WHERE* messages are used for node search.

The message format and operation coding is shown in Def. 1. The sequence number is required to detect the reception of multiple copies of the same message, a prerequisite for deployment of the Bluetooth device back-end that sends a message multiple times, explained in Sec. 4.2. The signature byte specifies the following tuple data pay-load. Depending on the back-end communication device and the supported packet format, the number of pay-load bytes can be very small. The signature field specifies the type of each tuple element with a tuple limit of four elements. For Bluetooth advertisement packages there is $N_{BLE}$=32, for the UDP back-end it is at most $N_{UDP} \geq 512$. The message header and the data payload is encoded in an BLE advertisement packet using one device local name attribute (ASCII85 encoded) and seven 16 Bit service UUID attribute fields, shown in Def. 1.

In contrast to typical tuple space services, the tuple operations are not atomic here. They can be executed at any given time point *t* in the near future or never, and the set of reachable tuple spaces that execute the request is not bound and can be zero. There is no assumption that neither a message arrives at a specific node nor that request is processed successfully. There are filter rules processed by agents that can be prevent tuple operation execution, too. That means, the *INP* operation is only a suggestion to all receiving tuple spaces to remove a matching tuple. All operations pose a probabilistic behaviour, i.e., there is a probability $\geq 0$ that a message is processed.

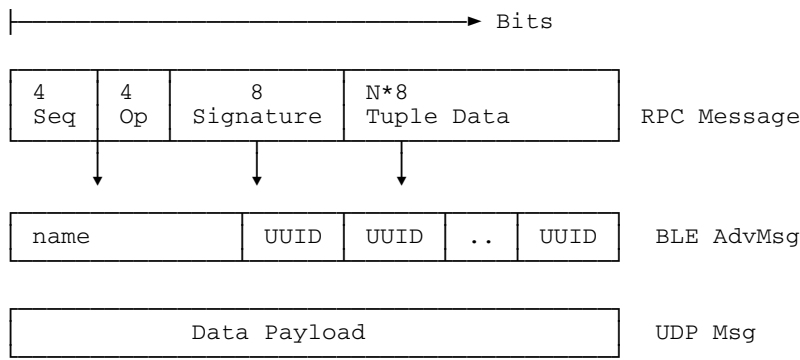

```
(2+N) Byte Messages (16 +N*8 bit)
                                    Bits

┌─────┬─────┬───────────┬───────────┐
│  4  │  4  │     8     │    N*8    │   RPC Message
│ Seq │ Op  │ Signature │ Tuple Data│
└─────┴─────┴───────────┴───────────┘

┌──────────┬──────┬──────┬────┬──────┐
│   name   │ UUID │ UUID │ .. │ UUID │   BLE AdvMsg
└──────────┴──────┴──────┴────┴──────┘

┌─────────────────────────────────────┐
│           Data Payload              │   UDP Msg
└─────────────────────────────────────┘

 Op = {
    0000  : 'IAMHERE',
    0001  : 'WHEREIS',
    0100  : 'OUT',
    0101  : 'INP',
    0110  : 'RD',
    0111  : 'TEST',
    1000  : 'TUPLE'
 }

Signature = [ TT , TT , TT , TT ]
```

```
TT = {
  00:   Formal Parameter,
  01:   String,
  10:   Int16,
  11:   Float32
}

String = [ C1,C2,.., CEND ]
Ci = ASCII 7 Bit
CEND = 0x80 & CharacterCode
```

*Def. 1. Tuple request message format, encoding in BLE/UDP packets, and operational codes*

The encoding of tuples is done automatically. Before a tuple is encoded and packed, a signature is derived, numbers are classified either in integer 16 Bit or float 32 Bit values depending on the actual value.

### 4.2 Bluetooth LE Back-end

Bluetooth is a standard radio communication technology connecting master with slave devices (peripherals) over short distances in the range of 10-300 cm. The data exchange is performed over negotiated connections. A connection is negotiated using the advertisement and scanning mode of Bluetooth devices. A peripheral device advertise its service by sending out short advertise message with preliminary information about the devices. A Bluetooth host (master) device, typically a smartphone, will receive these advertisement message if it is in scanning mode. One limiting factor of communication by mobile devices is power consumption, which is addressed by Bluetooth Low Energy (BLE) devices that can adapt to various communication situations with different power levels.

Devices can access remote tuple spaces of nearby neighbouring nodes (typicall in the range of 1-10m) by using BLE broadcasting (called ble-ts).

A device in advertisement mode will send out periodically advertisement message that contain a small payload depending on the advertisement message class, shown in Fig. 1. In this work, the pay-load is limited to 32 Bytes. There are 40 RF channels in BLE, each separated by 2 MHz (centre-to-centre). Three of these channels are called the primary advertising channels (labelled 37, 38, and 39), while the remaining 37 channels are called the secondary advertisement channels (they are also the ones used for data transfer during a connection) [10]. The primary channels are switched randomly in periods. On the other side, the scanning devices has to switch the (primary) receiving channels randomly, too. There is a probability $p$, that an advertisement packet is received if both scanner and advertiser are switch on the same channel and if there is no other sending within the receiving range creating collision (invalidation of the message).

The bandwidth and latency is limited by the advertisement time $t_{ad}=t_e-t_s$. The maximal number of independent messages that can be sent per second is $1/t_{adv}$. Assuming a channel switching time of $t_{sw} \approx 100$ms (a typical default value), a switching dead time $t_{de} \approx 2$ms, than $t_{ad} \geq 3t_{sw}$, and typically $t_{adv}$=500ms. The likelihood $p(N_r \geq 1)$ that a receiver $b$ receives a message from $a$ (i.e., at least one advertisement packet was received) depends on the distance $r_{a,b}$, the send power $P_t$, the antenna gains $G_t$ and $G_r$, the channel switching times $T_{sw}$ of $a$ and $b$, the receiver and sender dead times $t_{de}$ and the total active advertisement time $t_{ad}$, and the packet send frequency $1/t_{sn}$.

Approximately, if we assume $t_{sw,a}=t_{sw,b}=t_{sn} < t_{ad}/3$, and $t_{de}=0$, then we can estimate $p(N_r \geq 1)$:

$$P_r = P_t G_t G_r \left(\frac{\lambda}{2\pi r}\right)^2$$
$$p(N_r \geq 1) = \begin{cases} 1 - \left(1 - \left(\frac{3}{9}\right)^n\right) & \text{if } P_r \geq P_0 \\ 0 & \text{if } P_r < P_0 \end{cases}, \quad (1)$$
$$n = \frac{t_{ad}}{t_{sn}}$$

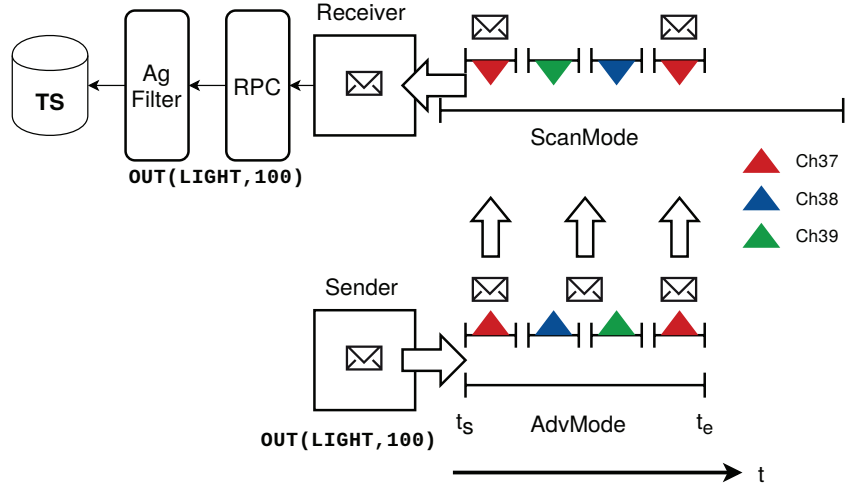

*Fig. 1. Principle BLE-based broadcast communication using advertisement packets. The sender and receiver switch their radio channels randomly and periodically. A packet m containing a tuple space request is sent multiple times in t ∈ [t$_s$,t$_e$].*

The simplified numerical computation of the reception probability $p_1$ is compared with simulations of two-node communication (still neglecting radio collisions) in Fig. 2 depending on the dead time of the receiver and sender of the BLE device. The computed results matches the simulation results approximately in the case of $t_{de} = 0$. An increasing dead time reduces the reception probability significantly. The simulation uses Monte Carlo simulation to introduce variance in the channel switching and packet sending.

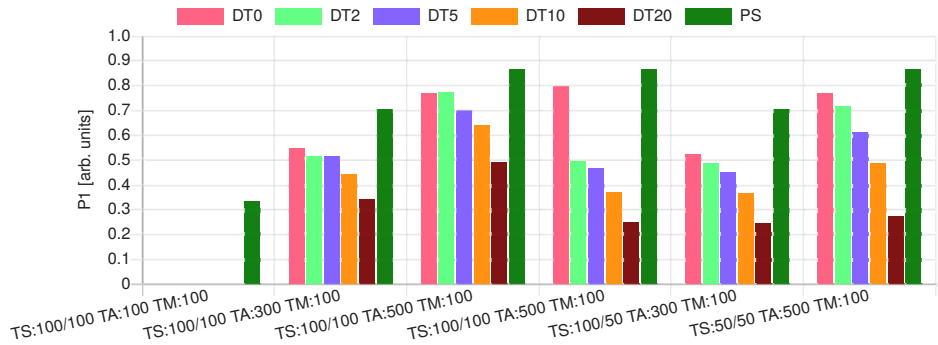

Fig. 2. Comparison of the .amth(p)$_1$ probability computed (PS) and calculated from simulation of two-node communication with different channel switching dead times (DT0=9ms, DT20?0ms, and so on)

### 4.3  UDP Back-end

In addition to the BLE broadcast communication, nodes that are connected to a local IP network can exchange tuple requests via UDP broadcast messages (called udp-ts). Although, UDP messaging is not reliable and there is no acknowledge of a packet delivery, the transmission probability for wired and switched connections is nearly 1 and mostly independent of the underlying network and the network load, in contrast, to the BLE-based communication with a transmission probability about 0.1-0.5 with a strong relationship to the radio communication load in the near neighbourhood of the stations. Wireless communication is different. Typically, WLAN do not support real broadcast communication, although it used broadcast radio, too. Experiments carried out in the use-case section show that wireless nodes can sent out broadcast UDP messages. These messages are always delivered to all LAN ports of the WLAN access point, but with a low probability below 0.5 to other wireless nodes. Today, all WLAN connection are secured an no other station can listen to the message stream (or at least decode them). Therefore, wireless broadcasting requires N:1 unicast messaging simulating broadcasting.

Initially, the UDP back-end sends a message only one time. Therefore, a much higher bandwidth and lower latency can be achieved. But experiments showed that some WLAN access points do not broadcast a received broadcast packet again to all connected client nodes. In dead, wireless UDP broadcasting is transformed in radio peer-to-peer multi-casting! The probability $p$ of a wireless connected device to receive UDP broadcast messages sent by other wireless connected nodes can be below 0.3. Therefore and optionally, the sending of UDP broadcast packets can be repeated like in the BLE back-end. The delta time is chosen randomly within a time interval [1,10] ms. Nodes connected wired via LAN do not show this packet dropping. All nodes connected wired to the WLAN access points receives all radio broadcasted UDP messages.

## 5. Security

On one hand, strong security in a non-negotiated and connectionless broadcast medium with message packet sizes below 64 Bytes is nearly impossible. On the other hand, this communication architecture and software framework should be deployed in public building and city environments controlling critical infrastructure, e.g., controlling ambient light intensity by user demands and user mobility. Any device can send tuple requests without any prior authorisation or authentication (Smartphones due this continuously). Today, billions of people using pandemic contact tracing Apps exactly doing this way and filling the air with broadcast advertisement messages (although, not all contact tracing applications rely on this methodology). So, the reception of broadcast messages cannot be prevented, and the major layer of security must be handled by the filter agents. Any time a message arrives from a sender there is a unique (MAC) identifier is annotated to each message. One approach is a list of authorised devices that are handled in groups by different agents. But the MAC identifiers must be transferred to all devices in the group in advance, which can be a show stopper for distributed applications of loosely coupled nodes.

A second approach uses symmetric two-way encryption with a private and a public key pair. The messages are encrypted using the private key (only known by trusted devices and users) and decrypted in the receiving device. But due to the hard data size constraints, only Format-preserving encryption (FPE) can be applied. Security and encryption is not addressed in this work. BeeTS implement a simple FPE algorithm that is able to encrypt and decrypt short data messages without compromising communication bandwidth or latency. The FPE algorithms can use any alphabet domain capable to encrypt both ASCII and binary data. It uses the aes-256-cbc algorithm. The encryption and decrpyion require each only about 0.05 μs/Byte on a typical desktop computer. Each tuple space can be protected with its own protection key and processing encrypted messages. An encryptor can be created on the fly (and used by agents, too). The encryptor is integrated in the BLE and UDP *rpc* modules before sending and after receiving a raw message. The encryption maps each data byte to an encrypted data byte independently using look-up tables, shown in Fig. 3. This kind of encryption is fast with low computational overhead, but is not safe against brute-force attacks. The mapping tables are created by using a user defined secret key.

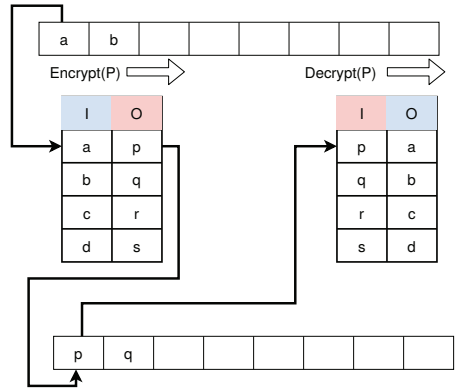

*Fig. 3. Format-preserving encryption of tuple messages using byte look-up tanles (P: Private secret password)*

## 6. BeeTS

### 6.1 Heterogeneous Networks

The principle network architecture combining Bluetooth and UDP-IP broadcast communication technologies is shown in Fig. 4. Tuple messages can either be sent via Bluetooth advertisement (based on [6]) or via single UDP packets within a local IP network. BeeTS is programmed entirely in JavaScript [11] and can be executed by node.js with a bluetooth socket modules for BLE access, the *noble* module for the central BLE part, and *bleno* for the peripheral part [12]. Note that BeeTS uses the peripheral and central (master) mode simultaneously (advertising and scanning), requiring a Bluetooth device with version ≥ 4.0. BeeTS is basically a small library module written in JavaScript.

Smartphones act as mobile devices and provide both a rich set of sensors and BLE connectivity. There are Cordova and a native Java App software. The Cordova App implements the original BeeTS library in a modified version to access BLE. UDP-based communication is actually not available for Smartphone versions. A hook solution for rapid prototyping uses a generic Smartphone browser with a suitable sensor Web API with a hybrid piggy-back architecture connecting the Smartphone (assuming it is Android-based) via USB to a Raspberry PI Zero computer. Using the *adb* tool a TCP port forwarding is established that enable HTTP-based micro service access of the Smartphone browser on the RP Zero. The micro service supports file sharing (HTML Web App + JavaScript code) and communication between the Smartphone and the RP Zero micro service. The RP Zero implements BeeTS, the Smartphone provides mobile sensor data. The sensor data is forwarded to the micro service by HTTP-JSON tuple space API access (RPC).

Each communication back-end can receive tuple requests. If there is a listener installed for tuples (with pattern matching), incoming tuples (*TUPLE* message) can be consumed by the listener or not. Otherwise, incoming tuples are stored in the local tuple space.

There are agents acting as a bridge between the communication back-end and the tuple space. They can filter incoming messages and decide to reply immediately, to access the tuple space, or to discard the message. Agents are functional code that listen to incoming tuple requests. There can be more than one agent. Communication between agents is established via the tuple space, too.

A broadcast message sending via BLE enables the advertisement mode of the device for a specific time interval $[t_s, t_e]$. The duration of the time interval $\Delta t$ determines the receiving probability, the collision probability (if more than one station is sending), the number of advertisement packets that contain the message $m$, and the number of different messages that can be sent per second. The interval time $\Delta t$ must at least $3 \times t_{sw}$, with $t_{sw}$ as the average channel switching time of the sender (and receiver). It is assumed that the sender and receiver have the same switching time, typically . Important to note that channel switching introduces small dead time intervals (about 1-10ms). A suitable value for $\Delta t$ is about 500ms.

Each physical communication interface (BLE/UDP) is attached to itws own tuple space, i.e., there ate two distributed space sets connected via BLE and UDP, respectively. This division is grounded in the spatial context of tuple spaces. Using BLE communication only nearby nodes can insert or remove tuples, whereas UDP communication enables tuple exchange over short and large distances, too. Tuple exchange between BLE and UDP tuple spaces is provided by a customisable router, shown in Fig. 4. Application-specific routing rules (functional code) provide transfers based on patterns and content of tuples. The rule set is dynamic and can be changed at run-time. The router extends the visibility and scope of tuples based on adaptive code. The code can use Machine Learning, e.g., reinforcement learning, to improve tuple space distribution. The routers connect local spaces and compose organised global spaces.

Each time a message is received it is passed to the Remote Procedure Call layer (RPC). Among the message data, the sender MAC ID, a time stamp, and the signal strength is added to the message.

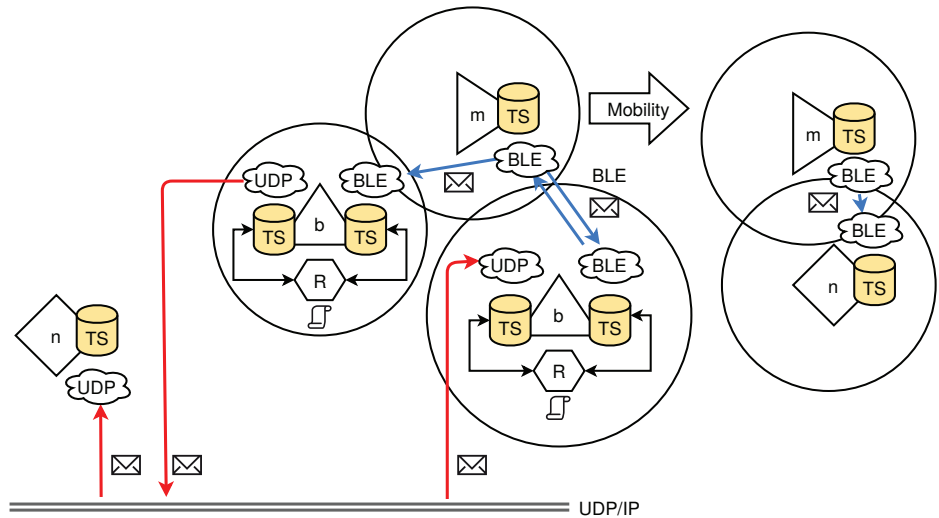

*Fig. 4. The hybrid network architecture using BLE and UDP-IP broadcast communication; n: stationary node, b: Stationary beacon, m: Mobile node, TS: Tuple space, R: Rule-based router*

### 6.2 Programming API

The programming API is rather simple. Examples are shown in Ex. 1. A tuple space is created for each communication back-end. For each device back-end there is the same set of operations that can be applied on the tuple space: *out* (persistent), *notify* (not persistent), *rd* (preserving), *inp* (destructive). All these operations create broadcast messages. To access the local tuple space directly, there is a mirror operational set accessed by the *host* attribute.

```
var ts1 = TS('udp');
ts1.start()
ts1.listen(['ALARM','LIGHT',null],function (t,source,rssi) {
  console.log('ALARM',t,source,rssi,t[2])
})
var ts2 = TS('ble',{keys:{private:'cloud1'}});
ts2.start()
ts2.listen(['SENSOR','LIGHT',null],function (t,source,rssi) {
  if (t[2]>500) ts1.out(['ALARM','LIGHT',t[2]]);
})
ts2.notify(['IAMHERE']);
ts2.rd(['SENSOR','LIGHT',null],function (t) { ... });
```

```
ts2.host.out(['SENSOR','LIGHT',1000]); // local op
ts2.out(['SENSOR','LIGHT',1000]); // remote op
```

*Ex. 1. BeeTS API examples*

The operations that have to wait for a reply always operating asynchronously with a callback function either called with the reply or on a timeout with empty data.

## 7. Agents

The BeeTS framework basically provides a communication platform using radio communication like Bluetooth or WLAN. The communication bandwidth of various devices can be significantly limited (e.g., in the case of Bluetooth advertisement mode that can be only 2 messages/second). One main feature of BeeTS nodes is the capability to execute event-based reactive agents programmed in JavaScript that perform, e.g., filtering of incoming tuple space requests, shown in Fig. 5. Agents can act as a bridge between different local tuple spaces, i.e., between ble-ts and udp-ts.

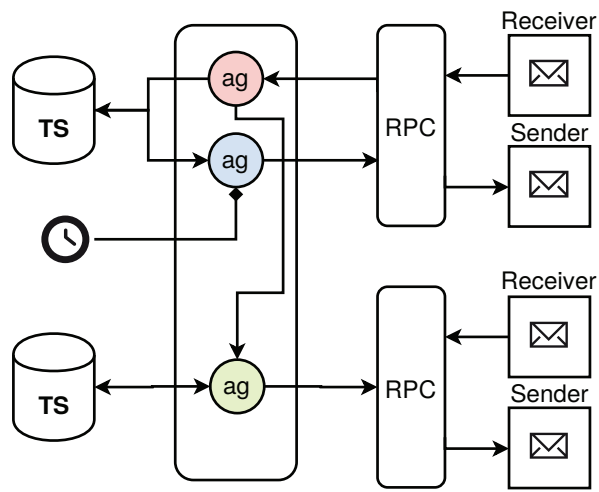

*Fig. 5. Agents create a bridge between tuple spaces with multiple communication devices (Red: Message event agent, Blue: Timer agent, Green: Router agent).*

An agent is functional data consisting of private body variables (including functional values) and event handlers that are activated by incoming messages, sensors (if the host platform provides them), periodically, or only one time. An event handler rule consists of an event expression that activates the rule and the hander function that is called on activation. An event rule activation expression can select three different event classes (*ts*, *sensor*, *timer*) with an optional conditional expression. Conditional expressions can access event source variables (actual sensor value, previous sensor value, tuple elements, time). Tuple listeners can consume tuples depending on the return value of the handler function. Not consumed tuples are stores in the local tuple space (if delivered by an *out* operation).

Agent event handler functions are application specific (see Ex. 2) and can be loaded remotely at run-time via a service API using active tuples. Agents can access the tuple spaces and communication interfaces programmatically as well as sending HTTP(S) requests to external services.

```
var agent = {
  x  : 0,
  y  : 0,
  ..
  on : {
    'ts.udp:(TIME,?)':  function (ev) {
      ts.ble.notify(ev.tuple);
      return consumed?;
    },
    'ts.ble:(SENSOR,?,?,?)':function (ev) {
      log(ev.tuple,ev.from,ev.rssi)
      ts.udp.out(['EVENT',JSON.stringify(ev.tuple),ev.from,ev.time]);
      return consumed?;
    },
    init : function () {
      this.x=random(1,1000);
    },
    1000 : function (counter) {
      // called each 1000 ms
      return true
    },
    'sensor.light:abs(sensor-sensor0)>50': function (ev) {
      if (ev.sensor>500) ts.ble.notify(['ALARM','LIGHT','HIGH']);
    }
  },
}
Agent.create(agent);
```

*Ex. 2. Example of an event agent with body variable and event handler sections*

As discussed in the next section, the tuple messages that can be sent via BLE have a strict size limit below 40 Bytes. Additionally, only a few independent tuple messages can be sent per second (typically 2/s). This requires progressive and tight scheduling of tuple messaging. This is a typical constraint solving problem that is performed by the multi-agent system, too. The agents have to satisfy the quality of service, e.g., a distributed human-interactive light managements in buildings (shown in the use-case section).

Agents are executed in sand-boxed containers. Since JavaScript objects and function can be serialised to text and deserialised at any time agent snapshots contain-ing actual agent data can be sent with active tuples to other nodes, i.e., tuples that contain code. Typically, encryption is used to secure agent migration.

## 8. Use-Case: Distributed Smart Building Control

This use-case deploys three different node classes implementing a distributed building light control system:

1. Stationary beacons (Raspberry 3) equipped with BLE and WLAN connectivity and supporting ble- and udp-conected tuple spaces in both test and production deployment;

2. Mobile devices (battery powered RP Zero stacked with a smartphone for testing, stand-alone smartphone in production systems) supporting ble- and udp-connected tuple spaces in test and ble-connected tuple spaces only in production environments;

3. A central monitoring and light control service supporting udp-connected tuple spaces.

The network architecture and experimental set-up is shown in Fig. 6. Each node deploys at least one event-based agent that implements necessary node operations like interaction with mobile devices or users, and tuple filtering and bridging. Beacons consume and aggregate mainly sensor data from mobile (sensorised) devices like smartphones and forward micro-surveys from the central server to mobile devices. But beacons can initiate and manage micro-surveys, too. To minimise the number of sent tuples via the BLE device, the mobile nodes monitor the user behaviour by analysing the accelerometer and gryoscope sensors. Updates of light sensor tuples are only sent if either the light conditions changes or the mobile device was moved in space. For rapid prototyping, smartphones are using generic Web browser loading an application page from the locally attached Raspberry PI zero bundled with the smartphone. All sensor data is sent to the embedded computer that executes the mobile application logic and that performed the BLE communication.

Mobile devices uses their light sensor in conjunction with accelerometer and gyroscope sensors to estimate the ambient light conditions and the user mobility by classifying the user activity in rest, smartphone use, and movement phases.

The measured light sensor data is processed by a sensor agent that tries to estimate if the smartphone is currently exposed to external light or if it is stored in a box. If external light is detected, sensor light tuples ⟨"SENSOR","LIGHT",value,time⟩ are sent via BLE. Nearby beacons distributed in the building about every 10-20m (and one per room/floor) collect these tuples and send aggregated sensor light values to the central server via udp-connected tuple spaces.

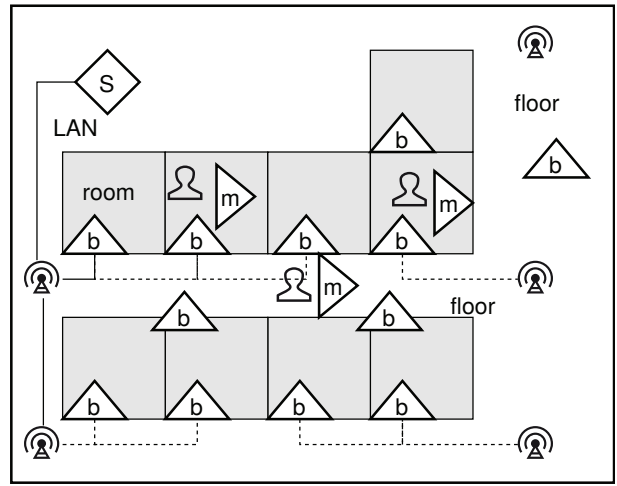

*Fig. 6. Experimental Set-up; s: Central server, b: beacon (Raspberry PI 3), m: Mobile device; Central server and beacons communicate via WLAN or LAN*

Among sensor tuples, there are micro-survey tuples that are sent from a beacon (initially delivered by the central server via the UDP tuple space) to mobile devices. If a device supports HMI (e.g., a smartphone), a short question is posted to a chat dialogue platform embedded either. The user can answer the question and the answer is passed back to the beacon (or any other beacon due to movement). The beacons collect the micro-survey replies and forward them to the central server.

The loss of tuple messages either due to out-of-reachability of beacons, or by collision, or by uncorrelated sender-receiver channel pairing, does not compromise the quality-of-service of the light control in the single rooms and floors. In average, 80% of the sent tuples were received and processed by the beacons. Mobile devices sent about 1 tuple message per second with an average minimal radio range availability of about 10 seconds (due to movement of the users, if any).

Using the light sensors, the mobility assessment using the accelerometer and gyroscope sensors of the mobile devices, and the performed micro-surveys providing user feed-back (satisfaction assessment), the illumination conditions could be optimised with respect to the user demands and energy consumption of about 30% without negative user feed-back and dissatisfaction.

For the evaluation of the loss rate of BLE tuple space communication, a partial set-up was chosen with four beacons at four different spatial positions and four mobile devices here all at the same position. The results of the measured average reception rate $R$ (loss is 100-$R$) is shown in Fig. 8. An average loss below 10% can be achieved within a radius of about 5m. Some nodes can communicate over larger distances up to 10m. The tuple message send time interval has no significant impact on the loss rate within time interval [500s,2000s] and with this (small) set-up. If the number of nodes within the radion range increases, the loss rate will increase.

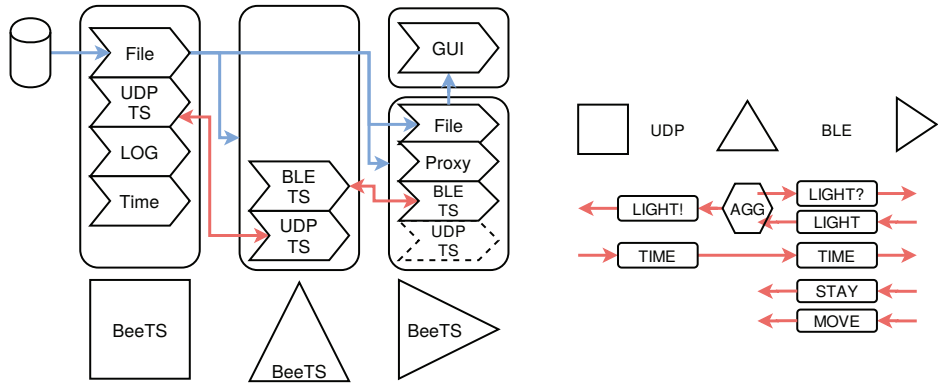

*Fig. 7. (Left) Communication and service layer architecture (Right) Tuple flows*

All nodes logged their tuple space activities via UDP broadcast communication accessing a tuple space on the central server. The server was connected via LAN to a WLAN acess point (AP), and all beacons and mobiles devices were connected to this AP. Broadcasting from LAN to WLAN did not work reliable (loss up to 50%), and even unicast UDP communication was not reliable via WLAN. This is a limitation of WLAN communication, although like BLE it is a broadcast medium, all device-AP communication is established as a peer-to-peer connection. A server has to simulate broadcasting by sending N:1 unicast messages.

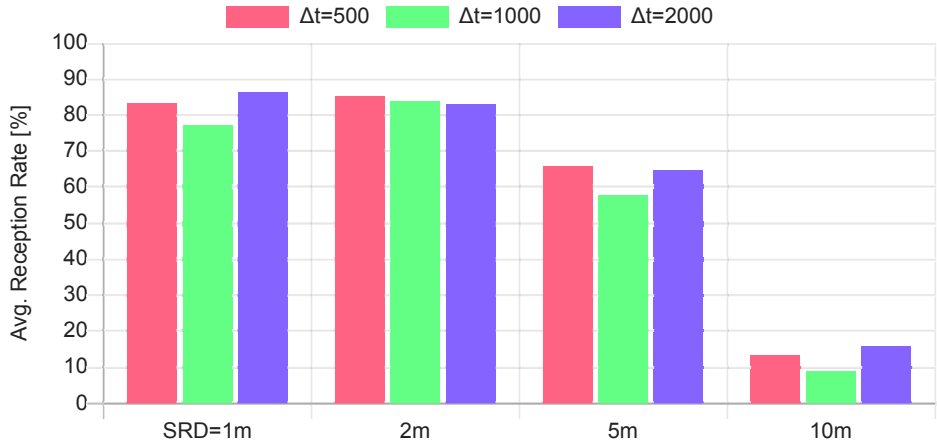

*Fig. 8. Average receptions rate of a sensor tuple (4 elements) in dependence of the distance between sender and receiver (SRD) and the tuple time interval $\Delta t$. Configuration: 4 beacons at different spatial positions and 4 mobile devices located at teh same spatial position*

## 9. Conclusion

In this work, distributed tuple spaces were used to exchange data between devices providing a spatial context. Smart devices access the tuple spaces by tuple message communication. The communication is connectionless and ad-hoc by using low-energy Bluetooth broadcasting available in any Smartphone and in most embedded computers, e.g., the Raspberry PI devices. Bi-directional connectionless communication is established via the advertisement and scanning modes by transferring encoded tuple messages. The communication nodes can exchange small data or functional tuples using the simple tuple space service API. Mobile devices act as tuple carriers that can carry tuples between different locations. Additionally, UDP-based Intranet communication can be used to connect tuple spaces. The Bluetooth Low Energy Tuple Space (BeeTS) service enables opportunistic, ad-hoc and loosely coupled device communication with spatial context. Multiple independent tuple spaces can be serviced on one network node. Among the tuple spaces, BeeTS implements simple reactive event-based agents. These agents perform tuple space management, interaction between devices and users, and they act as tuple filters and forwarders between multiple tuple spaces. A use-case study showed the suitability of the broadcast communication for distributed ad-hoc networks preserving a spatial context lacking in other approaches. Analysis showed a higher and unpredictable loss rate of UDP broadcast communication using WLAN, even if N:1 unicast communi-cation was used to simulate broadcast messages,

## 10. References

[1]     C. Maguire, *Attendance Tracking using Bluetooth Low Energy-Enabled Smartphones*, 2018

[2]     M. Cunche, A. Boutet, C. Castelluccia, C. Lauradoux, and V. Roca, *On using Bluetooth-Low-Energy for contact tracing*, Report, Inria Grenoble Rhône-Alpes; INSA de Lyon. 2020. hal-02878346v5

[3]     Reichert, L., Brack, S., & Scheuermann, B. (2020), *A Survey of Automatic Contact Tracing Approaches Using Bluetooth Low Energy*, Cryptology ePrint Archive

[4]     Li, J., & Guo, X. (2020), *COVID-19 contact-tracing apps: A survey on the global deployment and challenges* arXiv preprint arXiv:2005.03599

[5]     B. Skočir, G. Papa, A. Biasizzo, *Multi-hop communication in Bluetooth Low Energy ad-hoc wireless sensor network*, Journal of Microelectronics, Electronic Components and Materials Vol. 48, No. 2(2018)

[6]     M. Nikodem and M. Bawiec, *Experimental Evaluation of Advertisement-Based Bluetooth Low Energy Communication*, Sensors, vol. 20, no. 107, 2020

[7]     Davies, N., Friday, A., Wade, S. P., & Blair, G. S. (1998) *L2imbo: A distributed systems platform for mobile computing. Mobile Networks and Applications* 3(2), 143-156.

[8]     S. Bosse, *Unified Distributed Sensor and Environmental Information Processing with Multi-Agent Systems: Models, Platforms, and Technological Aspects*, ISBN 9783746752228, epubli, 2018

[9]     A. Bröring, A. Remke, and D. Lasnia, *SenseBox – A Generic Sensor Platform for the Web of Things*, 2012.

[10]    https://www.novelbits.io/bluetooth-low-energy-advertisements-part-1, accessed 1.2.2022

[11]    http://github/bsLab/beets, accessed 1.2.2022

[12]    https://github.com/noble/bleno, on-line, accessed 1.2.2022